\documentclass[cits]{PoS}

\title{Sunset integrals at finite volume}

\ShortTitle{Sunset integrals at finite volume}

\author{\speaker{Johan Bijnens}%
        \\
Deaprtment of Astronomy and Theoretical Physics, Lund University,\\
S\"olvegatan 14A, SE 22362 Lund, Sweden\\
        E-mail: \email{bijnens@thep.lu.se}}

\abstract{
Chiral Perturbation Theory is a useful tool to aid in performing the various
extrapolations needed in lattice QCD calculations of physical quantities.
These include extrapolations in quark mass, finite lattice spacing and
finite size of the lattice. Especially the latter will become more important
when the quark masses on the lattice become smaller. 

Here we develop the needed two-loop integrals at
finite volume to do the calculations for masses and decay constants for
all general mass cases.

I will present results based on an expansion in Bessel functions as well as
on a version using theta functions and compare their efficiency.
Work is in progress to combine these results with two-loop ChPT calculations.
}

\renewcommand{\logo}
{\raisebox{-5mm}
{\parbox{\textwidth}{\begin{flushright}
LU TP 13-35\\
October 2013
\end{flushright}
\includegraphics{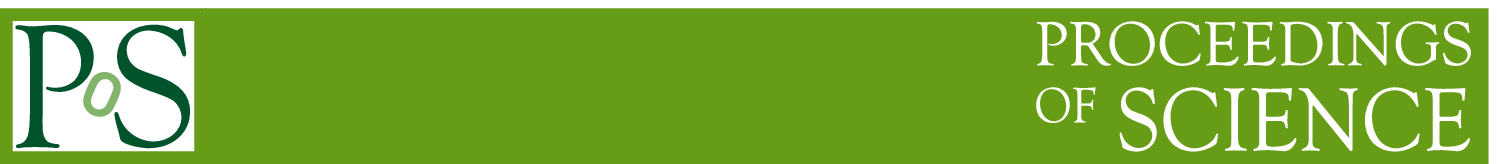}}
}}
\FullConference{{\rm To be published in the proceedings of:\\}
31st International Symposium on Lattice Field Theory
                 LATTICE 2013\\
		 July 29 -- August 3, 2013\\
		 Mainz, Germany}

\newcommand{\onloop}[1]{\lfloor #1\rfloor}

\newcommand{\sunset}[1]{\langle\langle #1\rangle\rangle}

\begin{document}

\section{Motivation}

Lattice QCD calculates at different quark masses and volumes and Chiral
Perturbation Theory has been very useful in the past for doing the
extrapolations in quark masses. Masses and decay constants are known
to two-loop order for the two- and three-flavour case as well as for all
mass cases with partial quenching
\cite{Lahde1,Lahde2,Lahde3}.
At finite volume as a pure ChPT calculation the mass for the two-flavour
case \cite{CH} and the vacuum expectation value in the three flavour
case is known \cite{BG}. The reason why we want to study also the general mass
case is that the Compton wavelength of the pion is about $1.4$~fm and
one might therefore have to go beyond the leading large volume $e^{-m_\pi L}$
terms with with present lattices. The convergence of Chiral Perturbation Theory
is governed by the resonance scale of $1/m_\rho\approx 0.25$~fm.

In the remainder I will work with an infinite extension in the time direction
and a size $L$ in the three spatial directions. The first work on finite volume
corrections in this context was done in \cite{GL2}. There is a large volume of
work at one-loop order but at two-loop only the above quoted exists.

I will first explain in detail the simplest one-loop case, tadpole integrals
and then introduce the extra parts needed to do the two-loop sunset
integrals. 

This work will be published in \cite{tobepublished}.
Some partial results\footnote{Note that there are misprints
in both \cite{Thesis} and my slides at the conference.} are in the master
thesis \cite{Thesis}.

\section{Underlying formulas}

In a finite volume the Fourier transform becomes a Fourier sum instead.
Let me first illustrate it in one dimension with periodic boundary
conditions $F(x+L) = F(x)$:
\begin{equation}
\int \frac{dp}{2\pi} F(p) \longrightarrow
\frac{1}{L}\sum_{p_n = 2\pi n/L} F(p_n) \equiv \int_L \frac{dp}{2\pi} F(p)\,.
\end{equation}
The integral with the subscript is defined to mean the sum. The problem is that
the sum is not simple to regulate when you have divergences. The Poisson
summation formula allows to again bring in an integral
\begin{equation}
\frac{1}{L}\sum_{p_n = 2\pi n/L} F(p_n)
=
\sum_{\ell=  n L}\int\frac{dp}{2\pi} e^{i\ell p}F(p)\,.
\end{equation}
If a twist angle $\theta$ is introduced in the boundary condition
$\theta$, $\phi(x+L) = e^{i\theta}\phi(x)$, we get instead
a sum over $p_n = (2\pi/L)\,n + (\theta/L)$ and
\begin{equation}
\frac{1}{L}\sum_{p_n = 2\pi n/L+\theta/L} F(p_n)
=
\sum_{\ell=  n L}\int\frac{dp}{2\pi} e^{i(\ell p-\ell(\theta/L))}F(p)\,.
\end{equation}

\section{One-loop tadpole}

Let me now illustrate the procedure on the simplest loop-integral, the one-loop
tadpole:
\begin{equation}
\onloop{X} = \int_V \frac{d^dr}{(2\pi)^d} \:
\frac{X}{{(r^2+m^2)}^{n}}\,.
\end{equation}
We do the Poisson trick in the three spatial dimensions
\begin{equation}
\onloop{X} = \sum_{l_r} \int \frac{d^dr}{(2\pi)^d} \:
\frac{X \: e^{il_r\cdot r-il_r \cdot \Theta}}{{(r^2+m^2)}^{n}},
\end{equation}
with $l_r=(0,n_1L,n_2L,n_3L)$, $\Theta=(0,\vec\theta/L)$. We split of the term
with $l_r=0$, i.e. the infinite volume term, and the rest
with $\onloop{X} = \onloop{X}^\infty_{} + \onloop{X}^V_{}$. The denominator
can be brought up with `$\alpha$' parameters,
$1/a = \int_0^\infty d\lambda e^{-\lambda a}$.
\begin{equation}
\onloop{1}^V = \frac{1}{\Gamma(n)}\sum_{l_r}^\prime \int \frac{d^dr}{(2\pi)^d} 
\int_0^\infty d\lambda \lambda^{n-1}\:
e^{il_r\cdot r-il_r \cdot \Theta} e^{-\lambda(r^2+m^2)}
\end{equation}
$\sum^\prime_{l_r}$ means sum without $l_r=0$ (all components zero).
We shift the integration momentum by $r=\bar r +i l_r/(2\lambda)$ to
obtain $
\onloop{1}^V = \frac{1}{\Gamma(n)}\sum_{l_r}^\prime
\int_0^\infty d\lambda \lambda^{n-1}\:
e^{-\lambda m^2-\frac{l_r^2}{4\lambda}-il_r\cdot\Theta} 
\int \frac{d^d\bar r}{(2\pi)^d}  e^{-\lambda \bar r^2}
$
This finally leads to the master formulae for tadpoles:
\begin{equation}
\onloop{1}^V = \frac{1}{(4\pi)^{d/2}\Gamma(n)}\sum_{l_r}^\prime
\int_0^\infty d\lambda \lambda^{n-\frac{d}{2}-1}\:
e^{-\lambda m^2-\frac{l_r^2}{4\lambda}-il_r\cdot\Theta}\,.
\end{equation}

We can now follow \cite{GL2} and do the integral over $\lambda$ and obtain
a sum over Bessel functions.
\begin{eqnarray}
\mathcal{K}_\nu^{}(Y,Z) &=& 
\int_0^\infty d\lambda \: \lambda^{\nu-1}_{} e^{-Z \lambda -Y/\lambda}_{}
= 2\left(\frac{Y}{Z}\right)^{\frac{\nu}{2}}_{}
K_\nu^{} \left(2\sqrt{YZ}\right),
\nonumber\\
\onloop{1}^V &=& \frac{1}{(4\pi)^{d/2}\Gamma(n)} \sum_{l_r}^\prime
e^{-il_r\cdot \Theta}
\mathcal{K}_{n-\frac{d}{2}-1}\left(m^2,\frac{l_r^2}{4}\right).
\label{tadpoleBessel}
\end{eqnarray}
This is valid also for noninteger dimensions $d=4-2\epsilon$
and can, if needed, be expanded in $\epsilon$.
The triple sum can be simplified: with
$\sum_{l_r}^\prime f(l_r^2) = \sum_{k>0} x(k)f(k)$, 
 $k=l_r^2$ and $x(k)$ is the number of times that $l_r^2=kL^2$.
The exponential decay for large $L$ follows directly from
$K_i(mL\sqrt{k}) \approx \sqrt{\frac{\pi}{2mL\sqrt{k}}}e^{-mL\sqrt{k}}$.

Alternatively, we can do the sum and obtain
an integral over Jacobi theta functions \cite{BV}.
The third Jacobi theta function is
$\theta_3(u|\tau) = \sum_n e^{i\pi\tau n^2+2\pi i u n}$. It satisfies
$\theta_3(u+n|\tau)=\theta_3(u|\tau)$
and $\theta_3(u|\tau) = \frac{1}{\sqrt{-i\tau}}e^{-\pi i\frac{u^2}{\tau}}
\theta_3\left(\left.\frac{u}{\tau}\right|\frac{-1}{\tau}\right)
$. Especially the latter is useful for small $\lambda$ for the tadpole integral
after doing the sum
\begin{equation}
\label{tadpoleTheta}
\onloop{1}^V = \frac{1}{(4\pi)^{d/2}\Gamma(n)}
\int_0^\infty d\lambda \lambda^{n-\frac{d}{2}-1}\:
e^{-\lambda m^2}
\Bigg[
\Pi_{j=x,y,z}\theta_{3}\left(-\theta_j/(2\pi)|iL^2/(4\pi\lambda)\right)
-1\Bigg]
\end{equation}
If no twist angles are present, it becomes a cubed theta function.
In the presence of twist angles, the trick with $x(k)$ of reducing the triple
sum to a single sum does not work, so (\ref{tadpoleTheta}) is usually the
better choice in that case.

\begin{figure}
\includegraphics[width=0.49\textwidth]{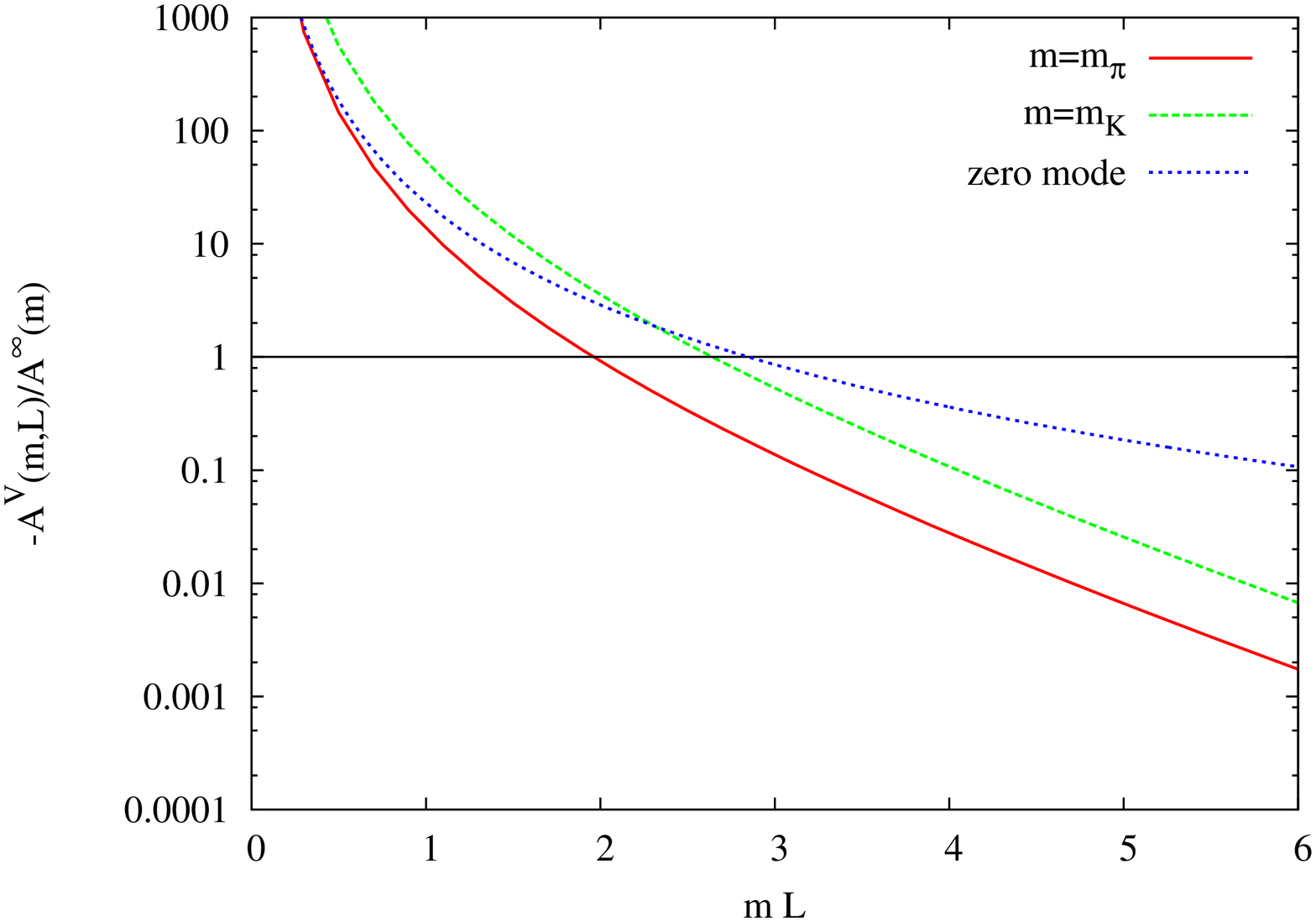}
~
\includegraphics[width=0.49\textwidth]{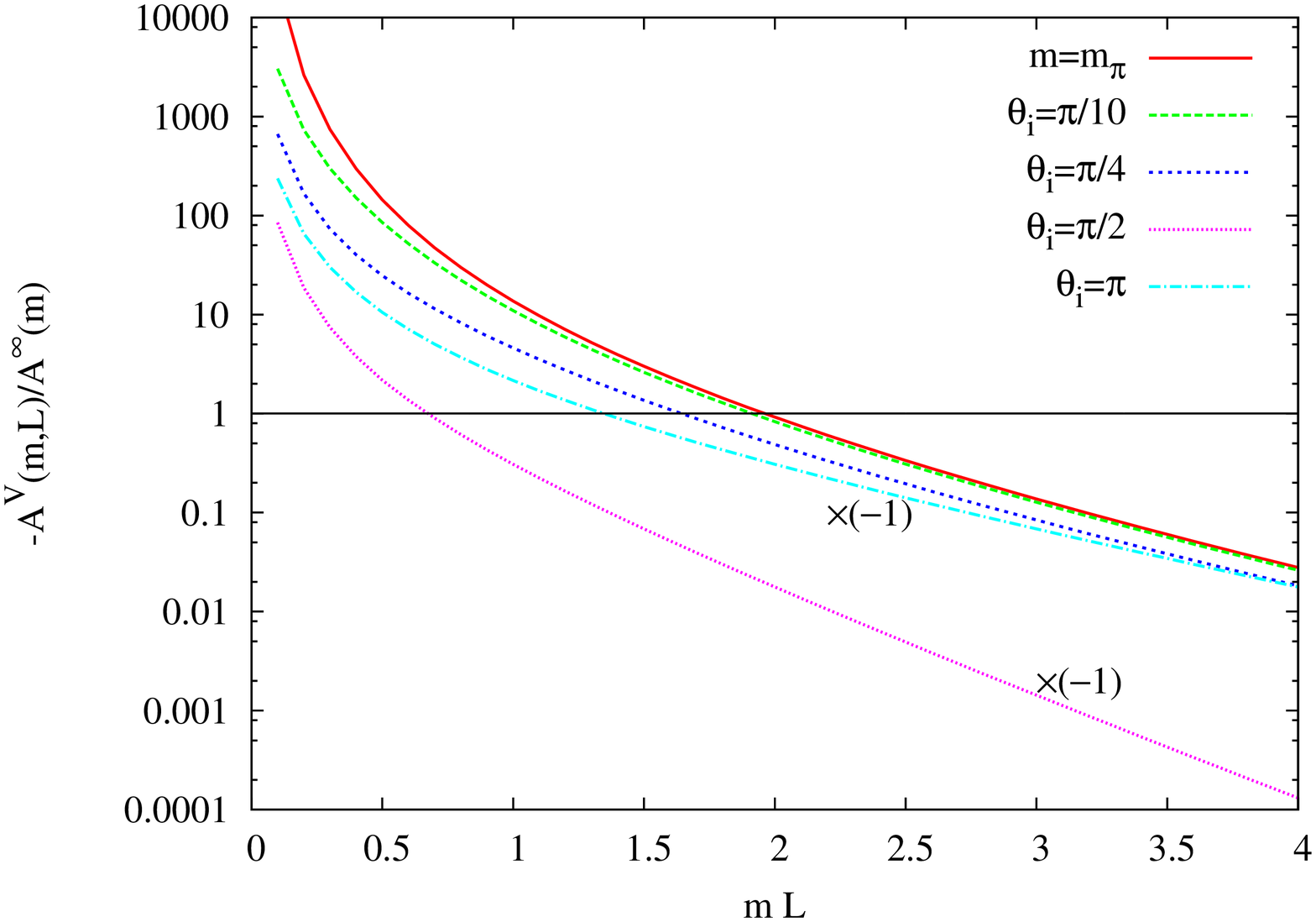}
\caption{\label{fig1} Left: The relative correction to the infinite volume
integral $A = \int_V \frac{d^dp}{(2\pi)^d}\frac{1}{p^2+m^2}$
for $\mu=0.77~GeV$ and $m=m_\pi,m_K$. Also show is the contribution from the
first spatial zero mode. Right:
The same but with a number of twist angles
$\theta_x=\theta_y=\theta_z=\theta_i$.
For $\theta=\pi/2$ all terms with an $(l_r)_i$ odd cancel, and the corrections
are much smaller.}
\end{figure}

As an example for the numerical size of the correction I show in Fig.~\ref{fig1}
the relative correction to the infinite volume integral
$A = \int_V \frac{d^dp}{(2\pi)^d}\frac{1}{p^2+m^2}$ for two different masses
as well as the result for some twist angles.

\section{More general one-loop integrals}

The methods above can be generalize to more complicated one-loop
integrals as long as one stays below thresholds. In particular, for
integrals with numerators $r_\mu r_\nu$ the steps are
\begin{itemize}
\setlength{\parskip}{0cm}
\item Shift with $r=\bar r+i l_r/(2\lambda)$
\item Integrals done with $\bar r_\mu \bar r_\nu\to \bar r^2 \delta_{\mu\nu}/d$
\item Similar for more complicated numerators
\item But extra terms show up: box and twisting break Lorentz invariance
\end{itemize}
For integrals with more denominators and external momentum:
\begin{itemize}
\setlength{\parskip}{0cm}
\item Combine the denominators with Feynman parameters
\item Shift with $r=\bar r+i l_r/(2\lambda)+(1-x)p$
\item This gives extra factors like $e^{-i(1-x)l_r\cdot p}$
\item Center of mass system $p=(p,0,0,0)$ $\Longrightarrow$ $l_r\cdot p=0$
\item Moving frame: deal with as for twist angle 
\end{itemize}
Because of the broken Lorentz invariance, there are more terms and this
will lead to extra form-factors at finite volume.

\section{Sunset integrals}

I will stick to the simplest sunset integral here. The more complicated
cases are treated in \cite{tobepublished}.
The simplest sunset integral is
\begin{equation}
\sunset{X} = \int_V 
\frac{d^dr}{(2\pi)^d} \frac{d^ds}{(2\pi)^d} \: \frac{X}
{{(r^2+m_1^2)}^{n_1^{}}{(s^2+m_2^2)}^{n_2^{}}{((r+s-p)^2+m_3^2)}^{n_3^{}}},.
\end{equation}
We now need to do the Poisson summation trick twice:
\begin{equation}
\sunset{X} = 
\sum_{l_r^{},l_s^{}} \int 
\frac{d^dr}{(2\pi)^d}
\frac{d^ds}{(2\pi)^d} \:
\frac{X \: e^{il_r^{} \cdot r}e^{il_s^{} \cdot s}}
{{(r^2+m_1^2)}^{n_1^{}}{(s^2+m_2^2)}^{n_2^{}}{((r+s-p)^2+m_3^2)}^{n_3^{}}}
\end{equation}
We stick here to the simplest case $X=1$, $n_1=n_2=n_3=1$ and $p=(p,0,0,0)$.
$l_r$ and $l_s$ are of the form $l_i=(0,n_1 L,n_2 L,n_3L)$.
For the tadpole case we split off $l_r=0$ but here it is a little
more complicated. First remove the infinite volume part with
\begin{equation}
\sunset{X} = \sunset{X}_{}^\infty + \sunset{X}_{}^V,
\end{equation}
But we also have the parts where only one of the loop momenta
feels the boundary or is quantized.
\begin{equation}
\sunset{1}^V = \sunset{1}_r^{}
+ \sunset{1}_s^{}
+ \sunset{1}_t^{}
+ \sunset{1}_{rs}^{},
\end{equation}
The parts are defined as
\begin{equation}
\sunset{1}_{\{{r},{s},{t},{rs}\}} =
\left\{
{ \sum_{l_r}^{\prime}},{ \sum_{l_s}^{\prime}},
{ \sum_{l_t}^{\prime}},{ \sum_{l_r,l_s}^{\prime\prime}} \right\}\times
\int 
\frac{d^dr}{(2\pi)^d}
\frac{d^ds}{(2\pi)^d} \:
\frac{\{{ e^{il_r^{} \cdot r}},{ e^{il_s^{} \cdot s}},
    { e^{il_t^{} \cdot (r+s)}},
    { e^{il_r^{} \cdot r}e^{il_s^{} \cdot s}}\}}
{(r^2+m_1^2)(s^2+m_2^2)((r+s-p)^2+m_3^2)}\,.
\end{equation}
The sums are over
$\{{ l_r\ne0,l_s=0};{ l_r=0,l_s\ne0};{ l_t\equiv l_r=l_s\ne0};
{ l_r\ne0,l_s\ne0,l_r\ne l_s}\}$.
The first three terms are those where only one-loop momentum $r,s$ or $r+s$
is quantized. The last one with both $r,s$ quantized.

\subsection{One momentum quantized}

We have sums over full momentum integrals, so we can use momentum redefinitions
to relate the three first terms:
\begin{itemize}
\item $(r\leftrightarrow s,l_r\leftrightarrow l_s)\Longrightarrow$
$\sunset{1}_s(m_1,m_2,m_3)=\sunset{1}_r(m_2,m_1,m_3)$
\item $(r\leftrightarrow t=r-s-p,l_t\leftrightarrow -l_r,e^{i l_t\cdot p}=1)\Longrightarrow$
$\sunset{1}_t(m_1,m_2,m_3)=\sunset{1}_r(m_3,m_2,m_1)$
\end{itemize}
So we only need $\sunset{1}_r$. For $\sunset{1}_r$
the $s$ integral is standard infinite volume:
\begin{equation}
\sunset{1}_r = \sum_{l_r}^\prime\int 
\frac{d^dr}{(2\pi)^d}
\frac{e^{il_r^{} \cdot r}}{(r^2+m_1^2)}
\int_0^1 dx
\frac{\Gamma\left(2-\frac{d}{2}\right)}{(4\pi)^{\frac{d}{2}}}\:
(\overline m^2)^{\frac{d}{2}-2}\,
\end{equation}
where we can use the usual expansions
$\frac{\Gamma\left(2-\frac{d}{2}\right)}{(4\pi)^{\frac{d}{2}}}\:
(\overline m^2)^{\frac{d}{2}-2} = \frac{1}{16\pi^2}\left[\lambda_0 
- 1 - \log(\overline m^2)\right] \:+\: \mathcal{O}(\epsilon)$ with
$\lambda_0^{} = 1/\epsilon + \log(4\pi) + 1 - \gamma$ and
$\overline m^2 = (1-x) m_2^2 + x m_3^2 + 
x(1-x)(r-p)^2$.

The part containing $\lambda_0$, $\sunset{1}_{r,A}$,
should cancel in the final result (of a physical quantity), so we ignore it.
We do partial integration in $x$ for the $\log(\overline m^2)$ term
obtaining
\begin{equation}
\sunset{1}_r = \sunset{1}_{r,A}+\sunset{1}_{r,G}+\sunset{1}_{r,H},\qquad
\sunset{1}_{r,G} = -\frac{1+\log(m_3^2)}{16\pi^2} \bar A^V(m_1^2)\,.
\end{equation}
The remaining part is
\begin{equation}
\sunset{1}_{r,H} =
\frac{1}{16\pi^2}\sum_{l_r}^\prime\int 
\frac{d^4r}{(2\pi)^4}
\frac{e^{il_r^{} \cdot r}}{(r^2+m_1^2)}
\int_0^1 dx x\frac{m_3^2-m_2^2+(1-2x)(r-p)^2}{\overline m^2}
\end{equation}
Bring up the denominators with `$\alpha$' parameters, shift $r$, do $\tilde r$
integral and finally symmetrize expression, and we get (details in \cite{tobepublished})
\begin{equation}
\sunset{1}_{r,H} =
\frac{1}{(16\pi^2)^2}\sum_{l_r}^\prime\int_0^\infty\! 
d\lambda_1 d\lambda_2d\lambda_3\frac{\lambda_3}{\tilde \lambda^2} e^{-M^2}
\left(m_3^2-m_2^2+\frac{\lambda_2-\lambda_3}{\tilde\lambda}
\left(2+\frac{\lambda_3+\lambda_2}{\tilde\lambda}
\tilde p^2
\right)\right)
\end{equation}
with
$
M^2 = \lambda_1 m_1^2+\lambda_2 m_2^2+\lambda_3 m_3^2
+\frac{\lambda_1 \lambda_2 \lambda_3}{\tilde\lambda} p^2
+\frac{\lambda_2+\lambda_3}{\tilde\lambda}\frac{l_r^2}{4}
$,
$
\tilde p = \frac{i l_r}{2}-\lambda_1 p,
$ and 
$
\tilde\lambda= \lambda_1\lambda_2+\lambda_2\lambda_3+\lambda_3\lambda_1
$. This contains a triple integral and a triple sum.

\subsection{Two momenta quantized: $\sunset{1}_{rs}$}

The same general method works:
Bring up denominators; Shift integration momenta; do the momentum integrals
and since it is finite, set $d=4$. This leads to the results
\begin{equation}
\sunset{1}_{rs}
=\frac{1}{(16\pi^2)^2}
\sum_{l_r,l_s}^{\prime\prime}\int_0^\infty\! d\lambda_1 d\lambda_2
d\lambda_3  \:
\tilde\lambda^{-2}e^{-\tilde M^2}
\end{equation}
with
$\tilde M^2 = 
\lambda_1 m_1^2+\lambda_2 m_2^2+\lambda_3 m_3^2
+\frac{\lambda_1 \lambda_2 \lambda_3}{\tilde\lambda} p^2
+\frac{\lambda_2}{\tilde\lambda}\frac{l_r^2}{4}
+\frac{\lambda_1}{\tilde\lambda}\frac{l_s^2}{4}
+\frac{\lambda_3}{\tilde\lambda}\frac{(l_r-l_s)^2}{4}$
and
$\tilde \lambda = \lambda_1\lambda_2+\lambda_2\lambda_3+\lambda_3\lambda_1$.
This contains a triple integral and a sextuple sum, very similar to
the previous subsection.

\subsection{Preliminary numerical results}

We basically proceed as in the tadpole case but with a few extras.
First we set
$\lambda_1=\lambda x$, $\lambda_2=\lambda y$, $\lambda_3=\lambda(1-x-y)$.
We can do the $\lambda$ integral and obtain a sextuple
sum over Bessel functions. This can be reduced to a triple
sum if there is no twist and we use that
$p\cdot l_r=p\cdot l_s=0$, since $p$ has no components
in the finite size directions. In that case
we have a triple sum over$k_1=l_r^2$,  $k_2=l_s^2$,
  $k_3=(l_r-l_s)^2$ and a quantity $x(k_1,k_2.k_3)$ that takes into account
how often each set of $k_i$ shows up in the sextuple sum.
Alternatively, the sum can be performed in 
terms of the Riemann or Siegel theta function:
\begin{equation}
\theta^{(g)}(z|\tau) =
\sum_{n \in \mathbb{Z}^g}
e^{2\pi i\left(\frac{1}{2}n^T \tau n+n^T z\right)}\,.
\end{equation}
This function has the useful properties
$\theta^{(g)}(z|\tau) = \theta^{(g)}(az|a\tau a^T)$ (with $a$ and $a^{-1}$ integer)
and 
$\theta^{(g)}(\tau^{-1}z|-\tau^{-1}) =
\sqrt{\det(-i\tau)}e^{\pi i z^T \tau^{-1} z}\theta^{(g)}(z|\tau)$. The latter
allows to speed up computation.

\begin{figure}
\includegraphics[width=0.49\textwidth]{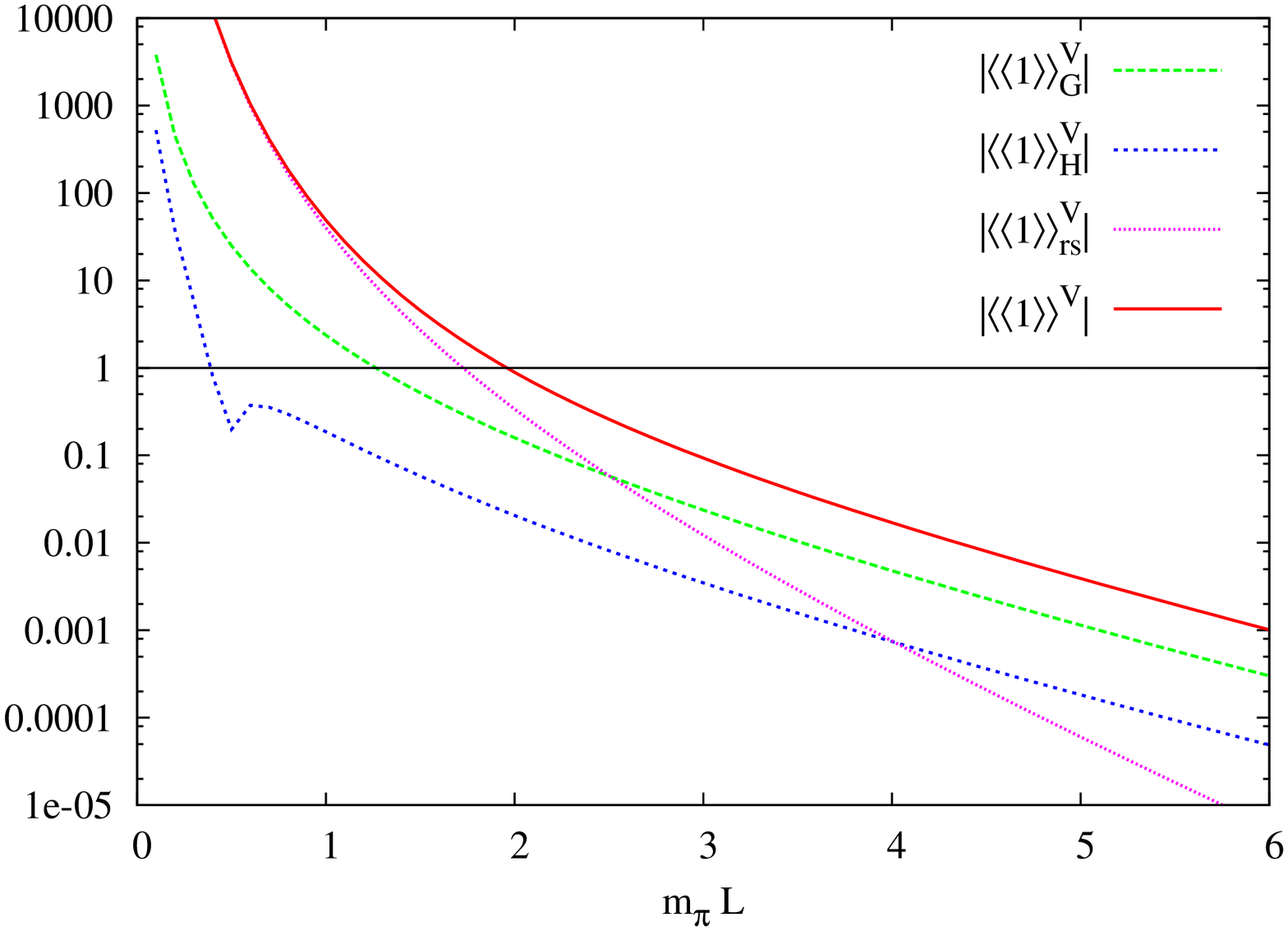}
~
\includegraphics[width=0.49\textwidth]{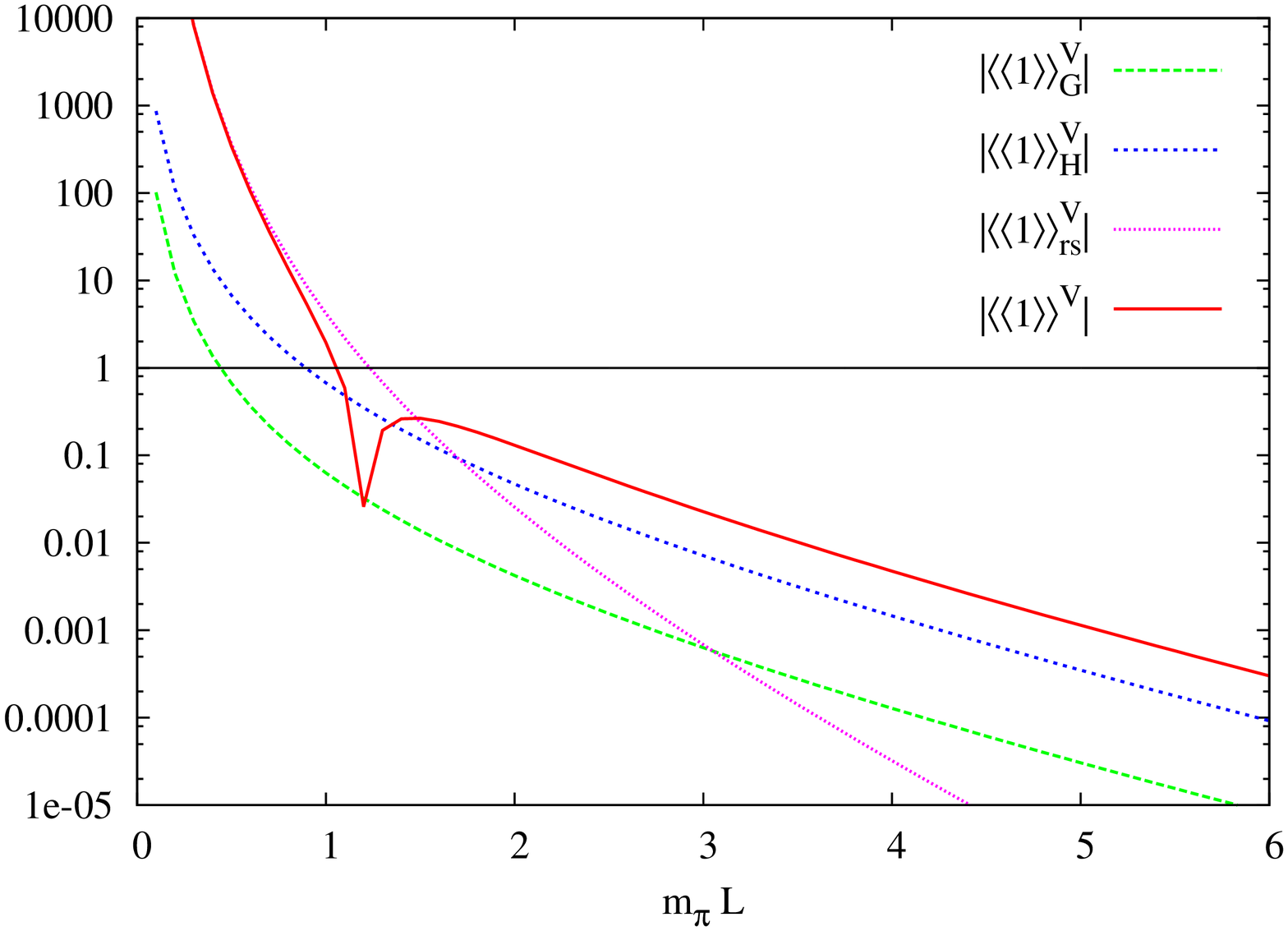}
\caption{\label{fig2} The simplest sunset integral and its various parts for
Left: $m_1=m_2=m_3=m_\pi$, $p^2=-m_\pi^2$ (pion on shell)
relative to $\sunset{1}^\infty = 3.7384~10^{-5}$~GeV$^2$.
Right:
$m_1=m_2=m_\pi$, $m_3=m_K$, $p^2=-m_K^2$ (kaon on shell)
relative to $\sunset{1}^\infty = 6.7407~10^{-5}$~GeV$^2$.
The funny bumps are not physical, we show absolute value
in a log plot and the result went through zero.}
\end{figure}

Some comments about the numerical work:
Getting 5-6 digits of precision for $\sunset{1}_{rs}$ goes fine,
it takes a while but is not too bad.
This method works below threshold.
For the speed; typically for large $mL$ the Bessel version is fastest, 
while for small or medium $mL$ the theta function works fastest.
The two methods always agree. Reaching a specified accuracy
is easier with the theta function version.
Some representative preliminary numerical results are shown in Fig.~\ref{fig2}.

\section{Conclusions and future}

An important hurdle in two-loop Chiral Perturbation Theory at finite volume
has been taken. Somewhat surprisingly, the various pieces in the sunset
are all needed dependent on the inputs used. A short side-note is
that the Riemann theta function in all its varieties not present in
mathematica (we need its derivatives) but needs to be programmed.

The cases with numerators are in progress \cite{tobepublished}.
Moving frame and/or twisting we have not studied yet, but I see no obvious
new problems appearing. 
The two-loop 3-flavour ChPT will need to be redone from scratch since
not all integral relations at infinite volume remain valid and they were
heavily used in the earlier work. This calculation is in progress.

\section*{Acknowledgments}

This work is supported in part by
the European Community SP4-Capacities
``Study of Strongly Interacting Matter''
(HadronPhysics3, Grant Agreement n. 283286),
the Swedish Research Council  grants 621-2011-5080 and 621-2010-3326.
I also thank my collaborators Emil Bostr\"om and Timo L\"ahde for a pleasant
collaboration.

\end{document}